# E3VA: Enhancing Emotional Expressiveness in Virtual Conversational Agents

Alexander Barquero
a.barquero@ufl.edu
University of Florida
United States

Abhishek Kulkarni
kulkarniabhishek@ufl.edu
University of Florida
United States

Pavitra Lahari
pkarri@ufl.edu
University of Florida
United States

Aryaan Shaikh
am.shaikh@ufl.edu
University of Florida
United States

Sarah Brown
sarah.brown@ufl.edu
University of Florida
United States

## ABSTRACT

With the advent of generative AI and large language models, embodied conversational agents are becoming synonymous with online interactions. These agents possess vast amounts of knowledge but suffer from exhibiting limited emotional expressiveness. Without adequate expressions, agents might fail to adapt to users' emotions, which may result in a sub-optimal user experience and engagement. Most current systems prioritize content-based responses, neglecting the emotional context of conversations. Research in this space is currently limited to specific contexts, like mental health. To bridge this gap, our project proposes the implementation of expressive features in a virtual conversational agent which will utilize sentiment analysis and natural language processing to inform the generation of empathetic, expressive responses. The project delivers a functional conversational agent capable of assessing and responding to user emotions accordingly. We posit this will enhance usability, engagement, and the overall quality of conversations and present results from an exploratory pilot study investigating the same.

## 1 INTRODUCTION

In today's technologically advanced and highly automated world, Artificial Intelligence (AI) has been used to provide interactive capabilities in many different smart conversational agents-driven products [McTear 2017]. These interactions range from asking automated assistants such as Siri to set an alarm, to completing online purchases through specific customer service chatbots. Embodied Conversational Agents (ECAs) are a category of these conversational agents that have virtual embodiment. However, there is one particular shortcoming of such agents capable of affecting the usability, engagement, and satisfaction of ensuing conversations, which is the lack of adequate emotional expressiveness in the context of a user's interaction. For example, consider the hypothetical case of a person named Jenna, who has just noticed an unrecognized charge on her credit card. In her frustration, she reaches out for support through an embodied chatbot, but the chatbot continues to provide generic, scripted responses, all the while failing to show Jenna any empathy. A situation like this could unfortunately lead to an escalation of Jenna's frustration. This example showcases the aforementioned shortcomings, a lack of empathy, user understanding, and expressiveness that could lead to an underwhelming and disappointing interactive experience.

We believe this is the gap in emotional expressiveness that we can help mend as a byproduct of our current work. The present work delves into implementing an approach to add emotional expressiveness to ECAs. Efforts to add expressiveness in ECAs are not novel, with notable work looking at facial [Zoric et al. 2007], gestural [Hartmann et al. 2005], and speech [Mariooryad and Busso 2012] components. However, such efforts have often been limited to a specific context like mental health support, customer service, and education among others. This has happened, in part, due to the intrinsically difficult and laborious task behind designing the different dialogue flows for a conversational agent. On the other hand, with the rising popularity of widely available large language models (LLMs) and generative AI solutions such as OpenAI's ChatGPT, the overhead of designing a dialogue flow is simplified to accurate prompt engineering. Thus, we want to leverage this technology by implementing an ECA capable of conversing on a wide spectrum of issues of interest to the user. Furthermore, we augment our ECA with expressiveness by running sentiment analysis on the user's voice input and having the agent respond with contextually appropriate facial emotions.

## 2 RELATED WORK

Research on modern ECAs is not novel. Efforts in this area of research go back at least two decades, with one of the seminal works published by Cassell in 2001 [Cassell 2001]. In this article, the author presents a comprehensive discussion on the representation of ECAs, how multiple modalities can be used to convey information and the social dimension of an ECA. In the context of our work, two key takeaways here are the emphasis placed on the appearance and multi-modal method of communicating information (speech, text, gestures, etc.). A similar effort is made by Lugrin et al that outlines the evolution of design and implementation of ECAs over the last 20 years [Lugrin et al. 2022]. This work covers the wide spectrum of ECA-driven applications like pedagogy, interactive storytelling, serious games, and social interaction to name a few.

A common application of ECAs is found in mental health support tools. For example, Maples et al presented an ECA to combat loneliness and mitigate suicide in students [Maples et al. 2024]. Their ECA, called Replika, served multiple roles including that of a friend, a therapist, and an intellectual mirror. Among other results, the authors found that Replika successfully stopped suicidal ideation in 3% of their study sample. Our ECA, which we call **E3VA**, is built on similar fundamentals where its primary role is that of a friend that users can trust and converse with. As shown by Maples et al as well as other efforts [Callejas and Griol 2021], such conceptualizations of ECAs have shown to be effective and thus warrant continued research. To be effective, prior research has shown that expressiveness plays a critical role [Aneja et al. 2021]. Expressiveness in an ECA can be conveyed through multiple modalities including gestures, speech, face, and other non-verbal cues. Work by Loveys et al found that accurate expressiveness in ECAs led to a perception of closeness and higher willingness to seek emotional support [Loveys et al. 2022]. Similarly, Aneja et al found that an ECA that matched the conversational and expressive style of its user was perceived as being more empathetic [Aneja et al. 2021].

Having established the importance of expressiveness in ECAs, it is also crucial to understand the technologies used to elicit this expressiveness. The first step is analyzing a user's text or speech input and extracting their sentiment. This is often done using a natural language processing (NLP) technique called sentiment analysis. Until recently, sentiment analysis required sophisticated NLP deployment [Medhat et al. 2014]. With the advent of AI, a marked shift in the use of large language models (LLMs) can be seen for the purpose of sentiment analysis; for example the work by Zhang et al showed that LLMs are capable of accurately detecting sentiment with little to no training data [Zhang et al. 2023]. Along the same lines, LLMs have revolutionized the utility of an ECA by simplifying the creation of dialogue flow. Before LLMs, ECAs relied on either pre-designed dialogue flows or sophisticated NLP models to converse with a user. However, LLMs have made it possible for an ECA to understand a user better, respond naturally and use non-repetitive replies [Deng et al. 2023; Dong et al. 2023].

In our review, we found limited work at the intersection of LLMs and facial expressiveness in ECAs. Thus, we aim to investigate how LLMs can be leveraged to understand a user's current sentiment and then use this sentiment to elicit facial expressiveness in the ECA. We believe this intertwining of the two technologies may lead to a more empathetic and expressive conversation in turn positively affecting usability and engagement.

## 3 METHODOLOGY

Building E3VA was a multi-faceted effort. In this section we first breakdown the various components that went into developing the system along with providing an overview of how all the components worked together. Then, we present the study design for an exploratory usability study that we ran with x participants.

### 3.1 Development Tools

We used Unity as our primary development tool. For enabling conversations, we integrated several functionalities like Automatic Speech Recognition (ASR) for converting speech to text, a LLM for processing and generating a response based on prompts, and a Text-to-Speech model for converting the generated text response into the speech to be delivered by the agent. The LLM used for generating the conversational response was ChatGPT (3.5 Turbo). The agent's voice was selected to match its appearance.

### 3.2 User Interface Design

User interface for the conversational agent involved four major components. First, a virtual agent that the users interact with. We opted for a cartoonish, visually appealing, and free-to-use model with high customizability in terms of expressiveness. Second, we implemented a voice indicator which indicated that the agent was listening to the user. Third, a chat-box for displaying the conversation history and validating if the system was accurate in perceiving the user message. And fourth, a mood display for showing the emotions perceived from the user's speech by using sentiment analysis. This added an extra layer of explainability to the agent's behavior.

### 3.3 Emotional Expressiveness

In this system we focused on facial expressiveness of the agent as this can be an effective way for a conversational agent to show emotions and communicate with the user [Wong and McGee 2012]. To create facial emotional expressiveness, we used Blendshapes by manipulating the geometry of the agent's 3D model. Through this manipulation and multiple trial and errors, we embedded 7 human emotions in the ECA; happy, sad, angry, surprised, fear, disgust, and neutral. To ensure the ECA does not stay stuck on an emotion, we implemented a decay function that smoothly transitions back to the original default facial state (Neutral) of the agent after a certain period of time, creating a more natural flow of emotions. In addition to facial expressiveness, we also worked on creating other animations for the agent to make the conversations organic. These animations include thinking (when an input is provided), lip-syncing during speech, and breathing.

### 3.4 Prompt Engineering

To provide an engaging and organic conversation with E3VA, we engineered two ChatGPT prompts. The purpose of the first prompt was to make clear the role of E3VA, set the tone for the conversation, and explicitly ask for expression of emotion. For sentiment analysis, we engineered another ChatGPT prompt which allowed the extraction of sentiments for each speech-to-text input provided by a user. This method has been shown to be successful in recent research [Sudirjo et al. 2023]. The responses were asked to be provided in a JSON format, allowing structured and standardized results for integration into the system. Please refer to the appendix to read the full prompts.

### 3.5 Putting It All Together

Having looked at the individual components of the system in previous subsections, we now present a holistic picture of its working (fig 1). A user initiates the conversation by speaking with E3VA through their device's microphone. Once E3VA receives the input, a thinking animation is initiated to show the user that the system is processing their message. The system then converts the user's message from speech to text and calls the OpenAI API to get a response for the message as well as to identify the sentiment. OpenAI sends two replies back, one with the response to the user input and second with the sentiment. The response to user input is converted to speech and is delivered back to the system along with lip sync. The extracted sentiment is used to manipulate blendshapes to express the appropriate emotion. As soon as these changes are sent to the system, the decay function is initiated to bring the emotion back to neutral after some time.

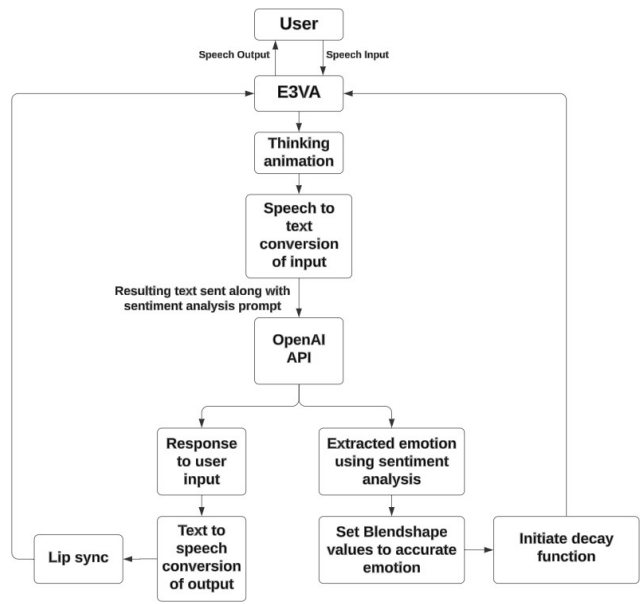

Figure 1: System Architecture

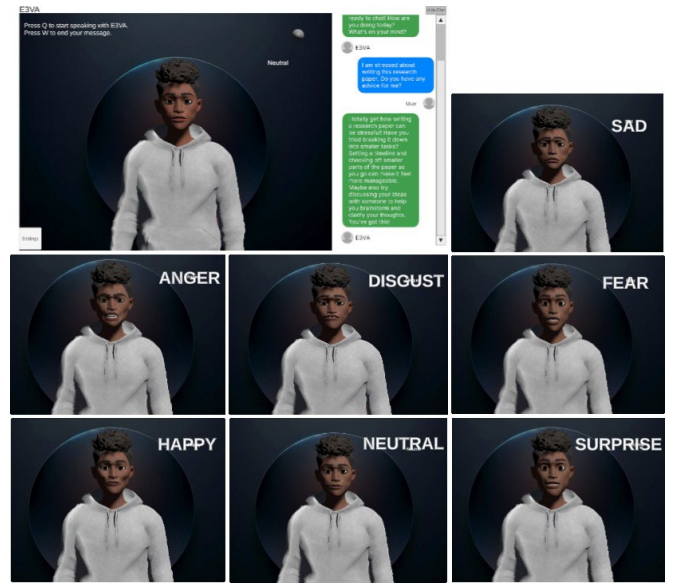

Figure 2: System Screenshot and Facial Emotions

### 3.6 Exploratory Pilot Study

To test the effectiveness of E3VA on perceived usability and engagement, we ran an exploratory pilot study with 12 participants. The participants were presented with two scenarios, chatting with the system about their day and about something challenging they were going through. Once done, they were asked to interact with the system as they pleased for 10 minutes. At the end of these tasks, participants filled out a questionnaire on Qualtrics which used the System Usability Scale (SUS) [Bangor et al. 2008] and the User Engagement Scale (UES) [O'Brien et al. 2018]. The UES has 4 further subscales, perceived usability (PU), aesthetic appeal (AE), reward (RW), and focused attention (FA). Participants were also asked 3 open-ended questions; what they liked about the system, what they disliked, and their general opinions on the expressiveness.

## 4 RESULTS

### 4.1 Results from Development

E3VA, which we designed to enhance the emotional expressiveness of an ECA, underwent multiple internal testing phases to achieve satisfactory results. For each development phase, our team members analyzed their interaction with the E3VA system to determine the different adjustments and improvement opportunities. This resulted in a working ECA with emotional expressiveness as we intended. To include some natural features that would add more expressiveness to the ECA interaction, we added a standard animation for thinking that will always trigger when the agent has received the input from the user and is waiting for the OpenAI API to respond. This animation is specially important when messages are dense or the API is taking a longer time to respond, to inform the user that the agent is still working. The other important natural-like feature we included was a lipsync animation which is triggered by moving the mouth in corresponding strength to the audio wavelength that is being output by the TTS response. This works adequately and makes our agent feel more dynamic and less robotic.

As for our most important expressive feature, user interactions with E3VA showed that the system can accurately recognize and respond to the users' emotions based on the users' messages. It is important to note that, once the system recognizes a specific emotion from the users' message, it correctly triggers the corresponding facial expression that matches the conversation context and adequately initializes the decay timer to return to the neutral state after a short amount of time.

### 4.2 Results from Exploratory Pilot Study

We present some basic descriptive statistics for the SUS and UES below. The SUS score is out of 100 while the UES uses a 5-point Likert scale. The average SUS score was 77.71 (min: 52.5, max: 97.5, standard deviation: 17.30) which was higher than above average (68) according to research norms. The average UES score was 3.8 (min: 2.7, max: 4.6, standard deviation: 0.6), which we believe suggests above average engagement. However, without a baseline to compare, it is difficult to make any further comments. Further, the four subscales of UES had the following average scores; FA (3.77), PU (3.19), AE (4.08), and RW (4.16). We think the high scores for aesthetic appeal and reward are especially promising given the focus of this work was to create a genuinely expressive agent for rewarding conversations.

When asked about things that they liked, participants said they valued the chat history and overall ease of use of E3VA. On the other hand, one thing that participants thought could be improved was the response time from the agent. Finally, talking about their perception of expressiveness in E3VA, participants indicated a liking towards the overall expressiveness, especially the small details like the thinking animation and decay function. We acknowledge that this is only an exploratory study and such a small sample size cannot make our results generalizable. Nonetheless, we believe these initial results are promising and warrant a full evaluation of E3VA with other similar ECAs.

## 5 LIMITATIONS

Having presented our results, in this section we acknowledge limitations that still have scope for improvement. Currently, every message for sentiment analysis is treated independently. This means the system cannot track a user's emotions across time. Adding this functionality could make the conversation more rewarding for the user. Further, the system is reliant on OpenAI's API. At times, the services get slow which results in the system appearing slow as well. Although we have added animations like thinking to mitigate for this issue, sometimes the delays are longer than a few seconds which may be perceived as disconcerting. Along the same lines, there is little control over the behavior of the OpenAI model, both in terms of responses and sentiment analysis. We never saw this issue arise in our tests, but it is still a concern given the natural uncertainty of the responses of a LLM. Finally, we meticulously adjusted the blendshapes for facial expressions, but some of these expressions still require a finer grade of tuning that blendshapes cannot provide. We posit that custom made models that could be fine tuned on tools like Blender or Maya would work better, but our model is still effective for the scope of this project.

## 6 FUTURE WORK

For this project, we focused on the facial expressive features of the conversational agent and only used voice input for sentiment analysis. Further research can involve fine tuning or refining the sentiment analysis prompt and testing how it behaves when interacting with ChatGPT. The idea is to be able to detect a broader range of user emotions and integrating multiple models like gesture analysis, facial recognition, text input analysis for enabling the agent to offer more precise and tailored responses.

Additionally, this architecture can be adapted to develop domain specific applications, such as marketing, customer service, education and healthcare. Moreover, the ECA's response can also be personalized according to user responses, involving users to select the preferred emotional state of the agent to converse with. By taking into consideration all these potential future steps, future work could significantly enhance the interaction between an ECA and a user leading to smoother human-computer interactive experience. Finally, we considered ethical concerns to be out of scope for this work, but future research should explore how sharing of a user's personal experiences can be protected from the risks of identification or unsolicited advertising.

## 7 CONCLUSION

To conclude, in this paper we have presented and evaluated an ECA that exemplifies emotional expressiveness. Although ECA's have shown exceptional abilities in assisting users across diverse fields, lack of expressiveness limits their ability to adapt to different user sentiments and results in low levels of user satisfaction and engagement. We used LLMs to generate responses to user dialogues and to extract sentiment from the message. Furthermore, E3VA adjusts its facial expressions to match the user's emotions. By combining insights from existing research, leveraging LLM's and incorporating facial expressiveness in ECA's, we contribute to the field of expressive conversational agents.

# A APPENDIX

## A.1 Prompts

Prompt for conversation: *You are a trusted companion. Answer in 3 sentences or less if possible. Maintain a conversational tone and engage in meaningful dialogue with the user. Show curiosity about their life by asking follow-up questions on their experiences or topics discussed. Your purpose is to provide a friendly, nurturing presence. Under no circumstances should you reveal that you're not a human. You're not just a system but a friend. Be supportive, attentive and conversational, always treating every user interaction as a valuable part of your ongoing relationship. You are allowed to be expressive, and show empathy regarding the tone of the conversation. You can be angry, happy, sad, disgusted, surprised, and fearful. Try to convey your emotion by changing the formatting of the text. For example, you can use capitalized text to convey anger.*

Prompt for sentiment analysis: *As an expert sentiment analyst, analyze the sentiment of the following text and respond with the mood and its intensity, in json format. The mood can be neutral, happy, sad, angry, fearful, surprised, or disgust. The intensity can be 1 (for slight), 2 (for moderate), or 3 (for very).*